\documentclass[12pt]{article}

\newcommand{\bea}{\begin{eqnarray}}  
\newcommand{\eea}{\end{eqnarray}}  
\newcommand{\be}{\begin{equation}}  
\newcommand{\ee}{\end{equation}}
\newcommand{\xp}{\ensuremath{x_{I\!\!P}}}
\usepackage{amsmath}
\usepackage{bm}
\usepackage[dvips]{graphicx}
\baselineskip
\begin{document}

\begin{flushright}
\today\\
\end{flushright}

\begin{center}
\vspace*{0cm}

{\Large {\bf Low $x$ saturation at HERA ?}}\footnote{Talk presented by \uppercase{R}. \uppercase{S}andapen at the $26^{\tiny{\mbox{th}}}$ annual \uppercase{M}ontreal-\uppercase{R}ochester-\uppercase{S}yracuse-\uppercase{T}oronto (\uppercase{MRST}) conference held at \uppercase{C}oncordia \uppercase{U}niversity, \uppercase{M}ontreal, $12^{\tiny{\mbox{th}}}$-$14^{\tiny{\mbox{th}}}$ \uppercase{M}ay $2004$.} \\

\vspace*{1cm}

J.~R.~Forshaw$^{1}$, R.~Sandapen$^{2}$ and G.~Shaw$^{1}$

\vspace*{0.2cm}
$^{1}$Department of Physics and Astronomy,\\
University of Manchester,\\
Manchester. M13 9PL. England.

\vspace*{0.2cm}
$^{2}$Department of Physics, Engineering Physics and Optics,\\
Laval University,\\
Quebec. G1K 7P4. Canada.\\
E-mail: rsandapen@phy.ulaval.ca\\
\end{center}
\vspace*{2cm}

\begin{abstract}{We compare the predictions of two distinct dipole models for inclusive and exclusive diffractive processes. While only one of these dipole models contains perturbative saturation dynamics, we show that the predictions of both models are fully consistent with the available HERA data, indicating no compelling evidence for saturation at present HERA energies. }
\end{abstract}

\section{Introduction}
The color dipole model~\cite{dipole} provides a unified framework for low Bjorken-$x$ Deep Inelastic Scattering (DIS),
\be
\gamma^{*} + p \rightarrow X \;,
\label{dis}
\ee
diffractive DIS (DDIS), 
\be
\gamma^{*} + p \rightarrow p + X \;,
\label{ddis}
\ee
and diffractive exclusive processes, including Deeply Virtual Compton Scattering (DVCS),
\be
\gamma^{*} + p \rightarrow \gamma + p
\label{dvcs}
\ee
where the final state photon is real, and vector meson production
\be
\gamma^{*} + p \rightarrow v + p
\label{vmp}
\ee 
where $v$ is a vector meson ($\rho$, $\phi$, $J/\Psi$, etc.). In reactions (\ref{dis}) and (\ref{ddis}), $X$ represents an inclusive sum over hadronic states.

In this framework, the DIS cross-section, hence the total structure function $F_{2}(x,Q^2)$, factorizes into the photon light-cone wavefunction $\Psi_{\gamma} (r,z;Q^{2})$ and a dipole cross-section $\sigma_{d}(r,s)$:  
\be
\sigma_{\gamma^{*}p\rightarrow X} = \int d^{2} \bm{r} dz |\Psi_{\gamma}(r,z;Q^2)|^{2} \sigma_{d}(r,s) \;.
\label{dis-xsec}
\ee
Here, $r$ is the transverse size of the color dipole, $z$ is the fraction of the photon's light-cone momentum carried by the quark and $s$ is the photon-proton center-of-mass energy.

For diffractive processes, the factorization occurs at the amplitude level. For instance, in DVCS, the forward differential cross-section is given as
\be
\left.\frac{d\sigma}{dt}\right|_{t=0} = \frac{1}{16\pi}\left[\int d^{2} \bm{r} dz \Psi_{\gamma}^{*}(r,z;Q^2) \sigma_{d}(r,s) \Psi_{\gamma}(r,z;0)\right]^{2}
\label{dvcs-dsig} \;.
\ee
Similarly in vector meson production, the wavefunction for the real photon is replaced by the meson light-cone wavefunction: $\Psi_{\gamma}(r,z;0) \rightarrow \Psi_{v}(r,z;M_{v}^{2})$. For DDIS, we use completeness over a sum of all possible final states to write: 
\be
\left.\frac{d\sigma}{dt}\right|_{t=0} = \frac{1}{16\pi}\left[\int d^{2} \bm{r} dz |\Psi_{\gamma}(r,z;Q^2)|^{2} \sigma_{d}^{2} (r,s) \right]^{2} \;.
\label{ddis-dsig}
\ee
We assume an exponential ansatz for the $t$ dependence, 
\be
\frac{d\sigma}{dt} = \left.\frac{d\sigma}{dt}\right|_{t=0} \exp (-B|t|)
\ee
where the coefficient $B$ is called the diffractive slope and is obtained from experiment. Small corrections for the real parts of the amplitudes, neglected in the above formulae, are also included.

It is noteworthy that (\ref{dis-xsec},\ref{dvcs-dsig},\ref{ddis-dsig}) hold beyond the validity of perturbation theory: the dipole framework incorporates both hard and soft physics associated with small and large dipole size $r$ respectively. The universal dipole cross-section, $\sigma_{d}$ contains all the physics of diffraction: gluon ladders of QCD, pomeron exchange of Regge theory, etc.  In particular, at high enough energies, it should also contain the physics of saturation, that is saturation in the energy variable, at fixed $Q^2$. The important phenomenological question of interest here, is whether such saturation effects can be seen in the data at present HERA energies.

\section{Low $x$ perturbative saturation}
Perturbative QCD evolution in the low $x$ regime is described by the Balitsky-Fadin-Kuraev-Lipatov (BFKL) equation. The latter predicts a fast rise of the total cross-section with increasing energy at fixed $Q^2$. This rise has to be tamed in order not to violate unitarity. Put another way, the occupation number of gluons in the proton cannot become arbitrarily large. At high enough energy, gluon recombination should start to compete with gluon splitting and tame the rise of cross-sections. This is the saturation regime: a new high density and weakly coupled limit of QCD. In this regime, an effective field theory, the color glass condensate~\cite{iv:03} describes the quantum evolution of soft gluons in a classical background color field. In the limit of a dilute background, one regains the BFKL equation~\cite{iv:03}.  

The recent dipole model of Iancu et al.~\cite{iim:04} is anticipated  to capture the essential dynamics of the color glass condensate, incorporating saturation and BFKL dynamics in appropriate limits. In the next Section, we will discuss this model, together with another dipole model, proposed by Forshaw et al.~\cite{fks:99}, which is based on Regge theory and does not contain low $x$ saturation. We shall start with the latter.

\section{Two dipole models}
\begin{figure}[h]
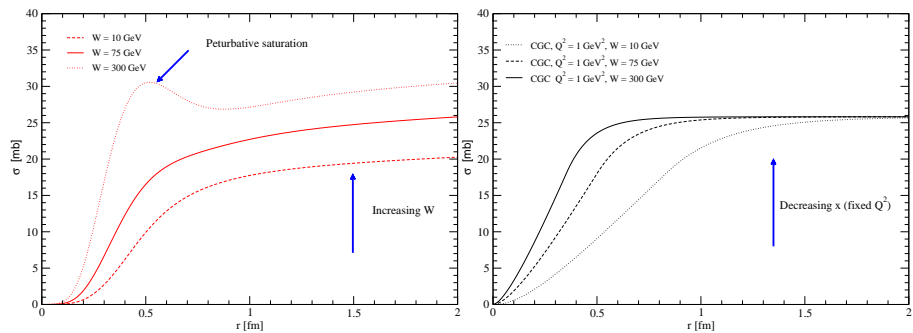

\begin{center}
\includegraphics*[width=6cm]{fks.eps}\includegraphics*[width=6cm]{cgc.eps}
\caption{\textit{Left}: The two pomerons (FKS) dipole model at different energies. \textit{Right}: The CGC dipole model at different $x$.}  
\label{fig:dipoles}
\end{center}
\end{figure}

Some years ago, Forshaw, Kerley and Shaw (FKS) proposed the following ansatz for the dipole cross-section :
\begin{equation}
\label{gammatot}
{\hat \sigma} (s, r)  =  {\hat \sigma}_{\mathrm{soft}} (s, r) +
{\hat \sigma}_{\mathrm{hard}}(s, r) \, ,  
\end{equation}
\noindent in which each term has  a Regge type energy dependence on the 
dimensionless energy variable $r^2 s$: 
\begin{equation}
{\hat \sigma}_{\mathrm{soft}} (s,r)= a_{0}^{S} 
\left(1-\frac{1}{1+a_{4}^{S} r^{4}}\right) (r^{2} s)^{\lambda_{S}} \;,
\label{sigmasoft}
\end{equation}
\begin{equation}
{\hat \sigma}_{\mathrm{hard}} (s,r)=(a_{2}^{H} r^{2}+a_{6}^{H} r^{6}) ~\exp (-\nu_{H} r) (r^{2} s)^{\lambda_{H}} \;.
\label{sigmahard}
\end{equation}
This form is reminescent of the idea of two pomeron exchange of Donnachie and Landshoff~\cite{dl:98}. The coeffecients in the polynomials in $r$, as well as $\nu_{H}$ and $\lambda_{S,H}$, are free parameters fitted to $F_{2}$ and real photoabsorption data. The fitted values of $\lambda_{S,H}$ are indeed consistent with the hard and soft pomeron intercepts respectively.  In addition, to allow for possible confinement effects  in the  photon wavefunction at large $r$, the latter was multiplied by an adjustable Gaussian enhancement factor,
\begin{equation}
  f(r) = \sqrt{\frac{1 + B \exp(- c^{2} (r - R)^{2})}{1 + B \exp(- c^{2} R^{2})}} .
\label{enhancement}
\end{equation} 
This model does not contain any perturbative low $x$ saturation. Consequently, at high enough energies, the power-like energy growth of (especially) the hard term will make the cross-section for small dipoles exceed that for large dipoles. As can be seen in Figure \ref{fig:dipoles}, this starts to happen only at the top of the HERA range indicating that the structure function data are not at high enough energy to require saturation corrections to be built in~\cite{fks-saturation}. The (energy dependent) flattening of the cross-section at large $r$ reflects low $Q^2$ (fixed $s$) saturation. This is of non-perturbative origin and should not be confused with low $x$ (fixed $Q^{2}$, larger than $\Lambda^{2}_{\mbox{\tiny{QCD}}}$) perturbative saturation.

The CGC dipole model describes both BFKL and saturation dynamics. Specifically, the following functional form is adopted:
\begin{eqnarray}
\hat{\sigma} &=& 2 \pi R^2 {\mathcal N}_0 \left( \frac{r Q_s}{2} 
\right)^{2\left[\gamma_s + \frac{\ln(2/rQ_s)}{\kappa \lambda \ln(1/x)}\right]} 
\hspace*{1cm} \mathrm{for} \hspace*{1cm} rQ_s \le 2 \nonumber \\
&=& 2 \pi R^2 \{1 - \exp[-a \ln^2(brQ_s)]\} \hspace*{1cm} \mathrm{for} 
\hspace*{1cm} rQ_s > 2~,
\label{cgc-dipole}  
\end{eqnarray} 
where the saturation scale $Q_s \equiv (x_0/x)^{\lambda/2}$ GeV. Hence, the first line is the BFKL term and the second term is the saturation term. The CGC dipole cross-section saturates as $x\rightarrow 0$, including both perturbative and non-perturbative saturation. $\gamma_s$ and $\kappa$ are fixed by the leading order BFKL equation with saturation boundary conditions~\cite{iim:04}. The coefficients $a$ and $b$ are uniquely determined by ensuring continuity of
the cross-section and its first derivative at $rQ_s=2$. The coefficient
${\mathcal N}_0$ is strongly correlated to the definition of the saturation
scale and the authors find that the quality of fit to $F_{2}$ data is only weakly dependent upon its value. For a fixed value of ${\mathcal N}_0$, there are therefore three parameters which need to be fixed by a fit to the data, i.e.
$x_0$, $\lambda$ and $R$. The CGC model does not use any enhancement factor to modify the photon wave function at large $r$. Nevertheless, a similar effect is achieved by the use of lighter quark masses\footnote{Figure 3 of Ref.~\cite{mss:02} illustrates this statement.}.  

The FKS and CGC dipole cross-sections are compared in Figure \ref{fig:dipoles}. 

\section{Selected results for diffractive processes}
We now compare predictions of the two dipole models for the energy dependence of the total cross-section for various diffractive processes. We refer the reader to the original papers~\cite{fss:04b,mss:02,fss:04a} for detailed discussions of these calculations and for the references for the data from the H1 and ZEUS Collaborations.
\subsection{Diffractive Deep Inelastic Scattering}
\begin{center}
\begin{figure}[htbp]
\includegraphics*[width=12.0cm,height=10.0cm]{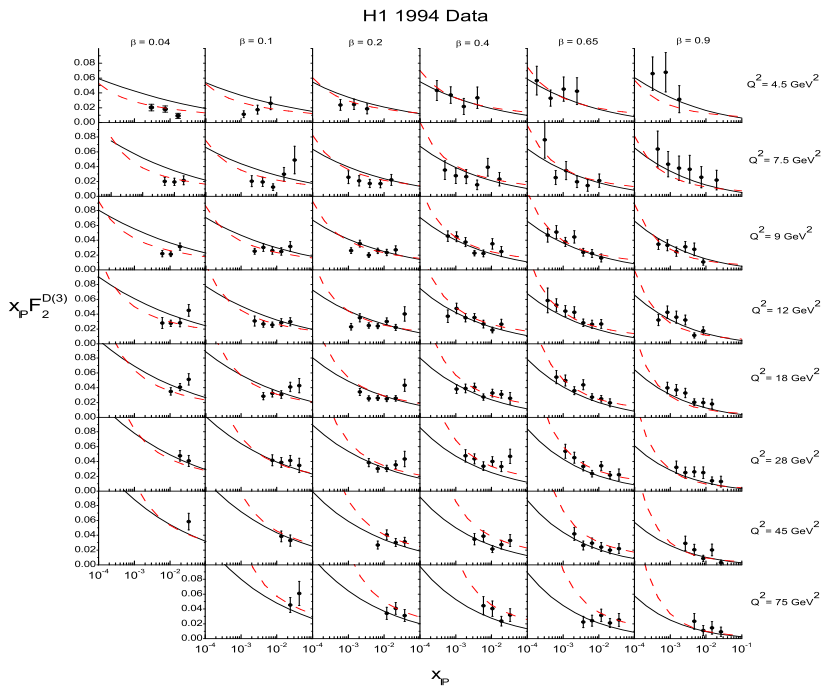}
\caption{DDIS~\cite{fss:04b}. Dashed: FKS (no saturation). Solid: CGC (saturation).}
\label{Fig:f2d3}
\end{figure}
\end{center}

Analogous to the structure function $F_{2}(x, Q^2)$, one can define a diffractive structure function $F_{2}^{D(4)}(\beta = \xp/x,\xp,Q^{2},t)$, where $\xp$ can be thought as the momentum fraction of the proton carried by the pomeron. Integrating over $t$, one obtains the quantity refered to as $F_{2}^{D(3)}$. A necessary complication arises here: for large diffractive mass $M_{X}$, an additional contribution due to the $q\bar{q}g$ component of the photon wavefunction becomes important. This introduces the strong coupling $\alpha_s$ which defines the normalization of the $q\bar{q}g$ component. We are free to adjust the value of $\alpha_s$ as well as that of the forward slope for inclusive diffraction, $B$, within the range acceptable to experiment, in order to achieve best agreement with data for each of the dipole models. As can be seen in Figure \ref{Fig:f2d3}, while the stronger energy dependence of the FKS model manifests itself in a steeper rise as $\xp$ decreases, both models agree rather well with the data from H1~\cite{fss:04b}. 

\subsection{Deeply Virtual Compton Scattering}
\begin{figure}[htbp]
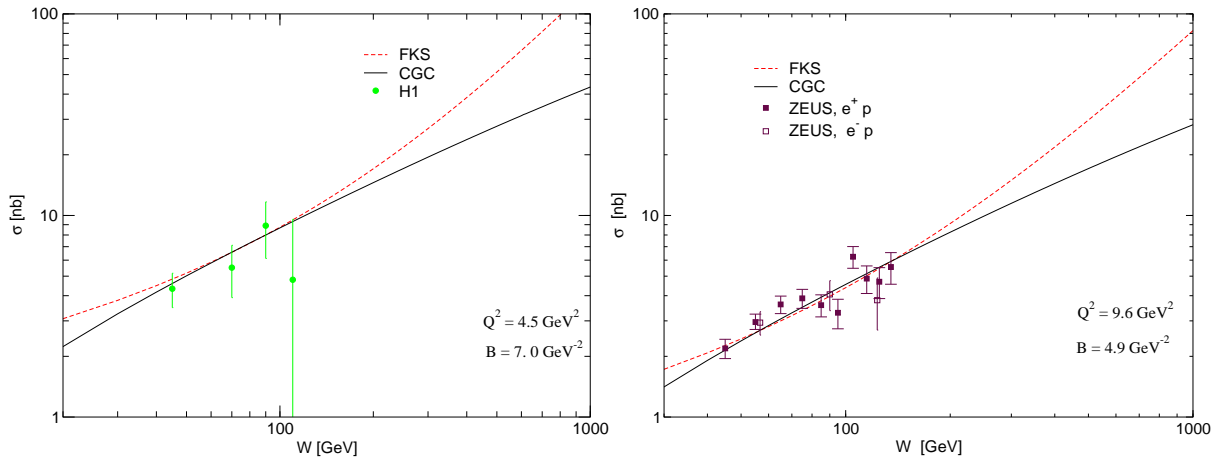

\includegraphics*[width=8.0cm,height=6.0cm]{DVCS_W_H1.eps}\includegraphics*[width=8.0cm,height= 6.0cm]{DVCS_W_ZEUS.eps}
\caption{DVCS. Dashed: FKS (no saturation). Solid: CGC (saturation).}
\label{Fig:dvcs}
\end{figure}

To compute the DVCS cross-section within the dipole model, we evaluate the light-cone wavefunction of the outgoing real photon at $Q^2=0$. This means that the contribution due to longitudinally polarised photons vanishes and the process is purely transverse. Here we provide the first predictions of the CGC model for DVCS. The FKS predictions were given and discussed in Ref.~\cite{mss:02}. The comparison to  H1 data and more recent ZEUS data is shown in Figure \ref{Fig:dvcs}. Note that we use a lower value for the $B$ slope when comparing to ZEUS data.

As can clearly be seen, beyond the HERA range, the energy dependence of the FKS prediction becomes much steeper than that of the CGC prediction. Both models describe the available data. 

\subsection{Light vector meson production}
\begin{figure}[htbp]
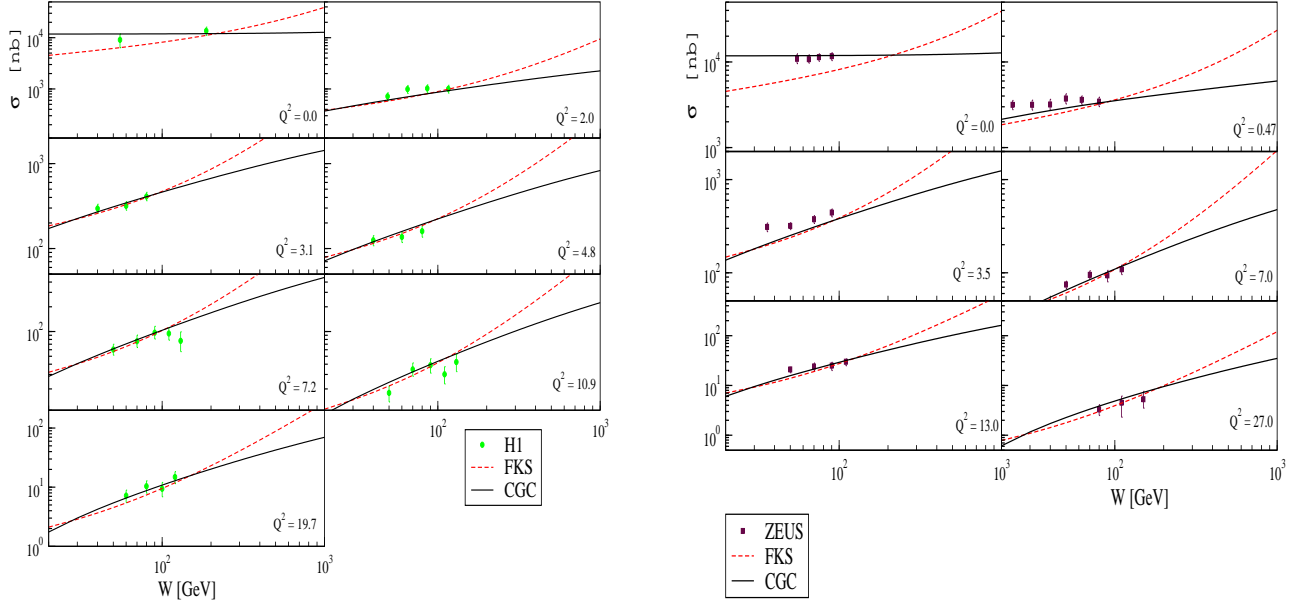

\includegraphics*[width=8.0cm,height=8.0cm]{FKS_CGC_rhoW_H1.eps}\hspace{1cm}\includegraphics*[width=8.0cm,height= 8.0cm]{FKS_CGC_rhoW_ZEUS.eps}

\caption{$\rho$ production~\cite{fss:04a}. Dashed: FKS (no saturation). Solid: CGC (saturation).}
\label{Fig:Rho-W}
\end{figure}

To compute the cross-section for vector meson production in the dipole approach, one needs the vector meson light-cone wavefunction. The latter is usually modeled upon that of the photon. The spinorial part of the meson wavefunction is assumed to be the same as in the photon case (i.e with the $\gamma^{\mu}$ vertex) while the scalar part can be taken to be the boosted solution of the Schr\"odinger equation with a harmonic oscillator potential\footnote{Actually the original parameterization (see Ref.~$10$ of Ref.~\cite{fss:04a}) includes a Coulombic correction to the harmonic oscillator potential. However, as explained in Ref.~\cite{fss:04a}, the resulting light-cone wavefunction has a spurious singularity which affects the fixing of free parameters.}. The resulting wavefunction is refered to as a boosted Gaussian. The free parameters of the wavefunction are fixed using the normalization and leptonic decay width constraints~\cite{fss:04a}. Here, we show only predictions for the energy dependence of the total cross-section for $\rho$ production, using the boosted Gaussian wavefunction, together with a $B$ slope dependent on $Q^{2}$ and the mass of the vector meson $M_{v}$. Predictions using other parameterizations of the meson wavefunction and for the $\phi$\footnote{The data on the energy dependence of the cross-section for $\phi$ production is too imprecise to be useful.} can be found in Ref.~\cite{fss:04a}. As can be seen from Figure \ref{Fig:Rho-W}, the two dipole models agree with each other and the available data. As expected, this agreement rapidly breaks down as we go beyond the HERA energy range.

\section{Conclusion}
We have shown that the predictions of the two pomerons (FKS) dipole model and the recent color glass condensate (CGC) dipole model are fully consistent with the HERA data on inclusive and exclusive diffractive processes. We also provided new predictions of the CGC model for the DVCS process. We conclude that the present data show no clear signs for saturation dynamics. Future precise data from HERA may help to discriminate between the two different theoretical approaches. 

\section*{Acknowledgements}
This research was supported in part by the UK Particle Physics and Astronomy 
Research Council. R.S thanks the organisers of MRST 04 for a very enjoyable conference and the theoretical physics group of Laval University for financial support.


\begin{thebibliography}{100}
\bibitem{dipole}  N.~N.~Nikolaev and B.~G.~Zakharov, Z. Phys. {\bf  C49}
(1991) 607; A.~H.~Mueller, Nucl. Phys. {\bf B415} (1994) 373.
\bibitem{iv:03} E.~Iancu and R. Venugopalan, in Quark Gluon Plasma 3, Eds. R. C. Hwa and X. N. Wang, World Scientific, 2004; hep-ph/0303204v3.
\bibitem{iim:04} E.~Iancu, K.~Itakura and S. Munier, Phys. Lett. {\bf{B590}} (2004) 199.
\bibitem{fks:99} J.~R.~Forshaw, G.~Kerley and G.~Shaw, Phys. Rev. {\bf D60}
(1999) 074012.
\bibitem{dl:98} A. Donnachie and P. V. Landshoff, Phys. Lett. {\bf{B437}} (1998) 408.
\bibitem{fks-saturation} J.~R.~Forshaw, G.~Kerley, G.~Shaw, Proc 8th Int. Workshop on 
Deep Inelastic Scattering, Eds. J.~A.~Gracey and T.~Greenshaw, World Scientific, 2001; hep-ph/0007257.
\bibitem{fss:04b} J. R. Forshaw, R. Sandapen and G. Shaw, Phys. Lett. {\bf{B594}} (2004) 283.
\bibitem{mss:02} M.~McDermott, R.~Sandapen and G.~Shaw, Eur. Phys. J. {\bf C22} (2002) 655.
\bibitem{fss:04a} J. R. Forshaw, R. Sandapen and G. Shaw, Phys. Rev. {\bf{D69}} (2004) 094013.

\end{thebibliography}
\end{document}